\documentclass[aps,preprint]{revtex4}

\newcommand{\ga}{\gamma}

\newcommand{\ph}{\phi}

\newcommand{\fr}{\frac}

\newcommand{\pe}{\perp}
\newcommand{\Pl}{\partial}

\newcommand{\ts}{\textstyle}

\newcommand{\bee}{\begin{equation}}
\newcommand{\ene}{\end{equation}}
\newcommand{\bem}{\begin{mathletters}}
\newcommand{\enm}{\end{mathletters}}
\newcommand{\bea}{\begin{array}}
\newcommand{\ena}{\end{array}}
\newcommand{\beea}{\begin{eqnarray}}
\newcommand{\enea}{\end{eqnarray}}

\newcommand{\bet}{\begin{tabbing}}
\newcommand{\ent}{\end{tabbing}}
\newcommand{\beb}{\begin{thebibliography}}
\newcommand{\enb}{\end{thebibliography}}

\newcommand{\fpar}[2]{\frac{{\ts \Pl \/ #1}}{{\ts \Pl \/ #2}}}

\newcommand{\npar}[3]{\frac{{\ts \Pl^{#1} \/ #2}}{{\ts \Pl \/ #3^{#1}}}}

\newcommand{\sech}{\mbox{sech$\,$}}

\begin{document}
\title{Fluid simulation studies of the dynamical behaviour of one dimensional 
relativistic electromagnetic solitons}
\author{ Vikrant Saxena, Amita Das, Abhijit Sen, Predhiman Kaw}
\address{Institute for Plasma Research, Bhat, Gandhinagar - 382428, India }

\begin{abstract}
\vspace{.5in} 
A numerical fluid simulation investigation of the temporal evolution of a special class of 
traveling wave solutions of the one dimensional relativistic cold plasma model is reported. 
The solutions consist of coupled electromagnetic and plasma waves in a solitary pulse shape 
({\em Phys. Rev. Lett. {\bf 68}, 3172 (1992)};{ \em Phys. Plasmas {\bf 9}, 1820 (2002)}). 
Issues pertaining to their stability, mutual collisional interactions and 
propagation in an inhomogeneous plasma medium are addressed.  It is found that solitary pulses
that consist of a single light peak trapped in a modulated density structure are 
long lived whereas structures with multiple peaks of trapped light develop an instability at 
the trailing edge. The interaction properties of two single peak structures show interesting 
dependencies on their relative amplitudes and propagation speeds and can be understood in 
terms of their propagation characteristics in an inhomogeneous plasma medium.

 \end{abstract}
\maketitle
\section{Introduction}
The  propagation and interaction of   ultra-high intensity lasers ($I \ge 10^{18} W/cm^2$ ) in plasmas   
display  a rich variety of nonlinear physics associated with the relativistic motion of the electron fluid 
in the electric fields of these lasers. An important early work in this area 
 is the pioneering study by 
Akhiezer and Polovin \cite{Akhiezer56} in $1956$. They
found exact nonlinear wave solutions for  
relativistically intense electromagnetic waves in a plasma that was modelled by 
coupled Maxwell and relativistic electron fluid equations. These early
theoretical results obtained well before the discovery of lasers have now become very relevant
due to the experimental realization of high intensity electric fields from modern powerful laser systems. 
This has promoted  a resurgence of activities in  this field. 
A variety of other nonlinear solutions, in the form of 
one dimensional nonlinear propagating solitonic structures have been obtained and 
discussed by various authors over the past several years 
\cite{decoster78,litvak79,esirkepov98,ksk,sudan97,kala1,kala2}.  One particular 
class of relativistic solitons, in the form of modulated light pulses coupled to plasma waves, has received 
special attention in the past decade in a number of analytic and simulation studies 
\cite{litvak79,esirkepov98,ksk,sudan97,kala1,kala2} and continues to be of current interest 
\cite{lontano03,farina05,berez05}. These solitary 
pulses are exact traveling wave solutions of the relativistic hydrodynamic equations and provide a useful paradigm 
for the investigation of various phenomena associated with intense laser plasma interactions that occur in 
applications like laser fusion, plasma based particle accelerators, photon acceleration schemes etc. In a 
strictly mathemetical sense these are not true soliton solutions, except in some simplifying limits when 
they obey canonical 
integrable model equations like the nonlinear Schrodinger equation. From a practical point of view it is 
therefore relevant to investigate such dynamical issues as the accessibility of these solutions, the 
stability and lifetime of these pulses, their mutual collision properties, their propagation characteristics 
in an inhomogeneous medium etc. Many of these issues are open questions today and are difficult to 
address analytically. In this paper we address some of these issues with the help of 
numerical simulation studies of the full set of 
relativistic one dimensional cold electron fluid equations along with the Maxwell's equations 
for the electromagnetic field evolution. The question of 
accessibility of these solitons was partially addressed in \cite{kim01,bulanov94,berez05} 
where the interaction of an 
externally imposed circularly polarized intense electromagnetic wave with a plasma was studied  
numerically by  particle as well as fluid simulations; and some part of 
of the laser energy was found to be trapped in non-propagating soliton 
like pulse structures. In our simulations we do not consider the vacuum plasma interface problem but confine 
ourselves to the dynamics of soliton solutions within the plasma. 
For this purpose we use the traveling wave solutions of 
\cite{ksk} as initial conditions and investigate their time evolution under various conditions. In general 
we find that these solitary pulses have a fairly long life time in that they continue to propagate stably 
through the plasma medium over several plasma periods. Single hump structures  are found to have longer life 
times than the multi-hump ones. 
The latter develop a distinct instability at the rear portion of the pulse and thereafter begin to radiate 
away. It is also observed that pure solitonic traits of preserving individual identities through a mutual 
collision process are displayed only by  small amplitude solutions obtained in the 
simplifying weak density  approximation. The general large amplitude solutions do not possess these solitonic traits. 
When these structures  propagate in an inhomogeneous medium they 
 gain (lose) speed while traveling down (up) the density gradient
and can be reflected back if the plasma density is large enough. In our present studies we have ignored ion motions and thermal effects. The influence of these effects 
will be studied in a future work. 

The paper is organized as follows.
The governing evolution equations are presented in the next section. We also briefly recapitulate here the 
solitonic traveling wave solutions of the relativistic cold fluid model as discussed in 
\cite{litvak79,ksk,esirkepov98,kala1}. In section III we
present results of our evolution studies for solitons propagating in a homogeneous plasma. We show 
that solitonic structures (both having single  and multiple peaks  of light waves ) survive with 
no distortion for several plasma periods. The single hump solutions are 
extremely  robust and have been observed to move undistorted for the entire simulation duration 
which was of the order of a few thousand plasma periods. The multiple peak solutions in contrast get 
distorted and develop an instability at their rear end  after around $30 - 40$ plasma periods.  

The interaction of two oppositely propagating structures 
are also considered in  section III. In section IV we investigate  the evolution of these 
solitonic structures in a plasma which is inhomogeneous i.e. for which the background plasma density 
varies spatially. Section V contains a summary and a 
discussion of our results. 
\section{Governing equations and stationary solutions}
 The  governing equations are the relativistic set of fluid evolution 
equations for a cold plasma in one dimension. We consider spatial variations to exist only along $x$, 
the direction of propagation, and consider the ions to be stationary. The relevant set of fluid and 
field equations are then, 
\bee
\fpar{n}{t} + \fpar{ (nu)}{x} = 0.
\label{d_cont} 
\ene
\bee
    (\fpar{}{t} + u \fpar{}{x})(\ga u) = \fpar{\ph}{x} - \fr{1}{2 \ga} \fpar{A_{\pe}^{2}}{x}
\label{l_mom} 
\ene
\bee
\npar{2}{\phi}{x} = n - n_0(x)
\label{pois}
\ene
\bee
   \npar{2}{\vec{A}_{\pe}}{x} - \npar{2}{\vec{A}_{\pe}}{t} =  \fr{n\vec{A}_{\pe}}{\ga} 
\label{wave}
\ene
where (\ref{d_cont}) is the electron continuity equation, (\ref{l_mom}) is the parallel electron momentum equation,
(\ref{pois}) is the Poisson's equation for the electrostatic potential $\phi$, (\ref{wave}) is the wave equation
for the vector potential $\vec{A}_{\perp}$ and other notations are standard. 
The perpendicular electron momentum equation has been integrated exactly to obtain the conservation of the transverse 
canonical momenta (sum of particle and the field momenta) as $u_{\perp} - \vec{A}_{\perp}/\gamma = 0$ and used to
eliminate $u_{\perp}$ in the above equations. 
Here $\ga$ is the relativistic factor 
$$
\ga = \sqrt{\fr{1 + A_{\pe}^{2}}{1 - u^{2}}}
$$
 In writing the above equations we have chosen to normalize 
the density by some appropriate density $n_{00}$. The length is normalized by the corresponding skin depth 
 $c/\omega_{pe0}$ (where $\omega_{pe0} = \sqrt{4\pi n_{00}e^2/m_e}$) and time by the inverse of the plasma frequency 
$\omega_{pe0}^{-1}$. The scalar and vector potentials are normalized by $mc^2/e$. In Poisson's equation 
$n_0(x)$ corresponds to the background ion density normalized by $n_{00}$.

The coupled set of nonlinear equations(\ref{d_cont}- \ref{wave}) 
 permit a variety of coherent solutions. A class of one dimensional propagating solutions with modulated 
 envelope structure 
of the above set have been obtained by using  the coordinate transformation $\xi = x - \beta t$ 
 and $\tau = t$ (where $\beta$ represents the group velocity of the structure). The vector 
 potential is assumed to be circularly polarized and has a sinusoidal phase variation of the form 
 $ \vec{A} = (a(\xi)/2)[\{\hat{y} + i\hat{z} \}\exp(-i \lambda \tau ) + c.c. ]$. The plasma
oscillations associated with the envelope structure are assumed to have no dependence on $\tau$. 
The above transformations convert Eqs.(\ref{d_cont},\ref{l_mom}) into ordinary differential equations 
which can be integrated to give $n(\beta - u) = \beta$ and $\gamma(1-\beta u) - \phi = 1$, where 
one assumes that at the boundaries $u = 0$, $\phi = 0$ and $n = 1$. One eliminates $n$ to write 
Poisson's equation (Eq.(\ref{pois})) as 
\bee
\phi^{\prime \prime} = \frac{u}{(\beta -u)}
\label{pois_st}
\ene
Here $^{\prime}$ denotes derivative with respect to $\xi$. 
Writing $a(\xi) = R exp(i \theta)$, the wave equation (Eq.(\ref{wave})) can be written as 
\bee
R^{\prime \prime} + \frac{R}{1-\beta^2} \left[\left(\lambda^2 -\frac{M^2}{R^4}\right) \frac{1}{1-\beta^2} 
- \frac{\beta}{\beta - u} \frac{1-\beta u}{1+\phi}\right] = 0
\label{wave_st}
\ene
Here $M = R^2[(1-\beta^2)\theta^{\prime}-\lambda \beta]$ is a constant of integration and $R^2 = A_y^2 +A_z^2$. 
Eqs.(\ref{pois_st},\ref{wave_st}) form a coupled set of second order differential equations 
in two fields $\phi$ and $R$ respectively. 
The longitudinal velocity $u$ appearing in the two equations 
can  be expressed entirely in terms of $R$ and $\phi$ as 
\bee
u = \frac{\beta(1+R^2) - (1+\phi)[(1+\phi)^2 - (1-\beta^2)(1+R^2)]^{1/2}}{(1+\phi)^2 + \beta^2(1+R^2)}
\label{u_st}
\ene
Eqs.(\ref{pois_st},\ref{wave_st}) have been solved by Kaw {\it et al.} \cite{ksk} and others 
\cite{kala1} for $M = 0$. The analytical solutions for the above equations 
were obtained only in the weak density response limit where the approximation $\phi \approx (1+R^2)^{1/2} - 1$
holds and the coupled set of 
Eqs.(\ref{pois_st},\ref{wave_st}) can be reduced to a single equation in $R$. 
The amplitude of the solutions in this case are small and retaining terms upto the cubic order in ($R$) 
the reduced equation becomes the nonlinear Schrodinger equation, viz.
\bee
R^{\prime \prime} - c R + d R^3 = 0 
\label{nls}                                                                 
\ene
where $c = (1-\beta^2 -\lambda^2)/(1-\beta^2)^2$ and $d = [4 \lambda^2 -(1-\beta^2)(3+\beta^2)]/[2(1-\beta^2)^3] $. 
Equation (\ref{nls}) can be integrated exactly to obtain 
a soliton solution of the $\sech$ form
\bee
R = R_m  \sech (\sqrt{c} \xi)
\label{lamp_sol}
\ene
where 
\bee
R_m =\sqrt{\left( \frac{2c}{d} \right)} = 2 \left[ \frac{(1-\beta^2)(1-\beta^2 -\lambda^2)}{4\lambda^2 - 
(1-\beta^2)(3+\beta^2)}\right]^{1/2}
\ene
The analytic form of the potential is then, $\phi \approx (1+R^2)^{1/2}-1 \approx R^2/2 = 
(R_m^2/2) \sech^2(\sqrt(c)\xi)$. 
It is thus clear that for low amplitude ($R_m << 1$) solutions, $\phi$ is invariably much smaller than $R$.

In the absence of any such simplifying assumption the equations (Eqs.(\ref{pois_st},\ref{wave_st})) 
cannot be solved analytically. However, for the general case several varieties of 
 numerical solutions have been obtained. The simplest variety is that  where the modulated structure 
 involves only a single peaked structure for both $\phi$ and $R$. Such a solution is displayed 
 in the top subplot of Fig.1. Another class of solutions involves multiple peaks for the radiation field $R$ trapped within 
 a modulated  envelope of scalar potential $\phi$. For these structures the amplitude of scalar potential 
 is very high compared to that of the radiation field $R$. We show one such solution in the bottom  
subplot of Fig.1 for the parameters 
 $\lambda = 0.44466066$ and $\beta = 0.5$. These two classes of solutions have different spectral properties.
 The single hump solitons have been found to have a continous spectrum in the $\lambda$ and $\beta$ parameter space whereas 
 the multi-hump solitons exist at discrete values of $\lambda$ for a given value of $\beta$. The spectral domains 
 of these solutions have been numerically obtained in \cite{kala1} and we reproduce the $\lambda - \beta$ diagram 
 of that paper as Fig.2 here. The parameter $p$ refers to the 
 number of extrema in the radiation field (for instance  multihump structure in Fig.1 would correspond to $p = 9$). 
 The uppermost curve in Fig.2  corresponds to the 
 weak density response limit where the analytical $\sech$ solutions given by Eq.(\ref{lamp_sol}) hold. 
 The shaded region shows the continuum values of $\lambda$ and $\beta$ for which general single hump ($p=1$)solutions 
 have been obtained. 
 As one goes down in $\lambda$ in the shaded region from the upper curve one moves away 
 from the weak density response analytical solutions to the large amplitude numerical solutions 
 of Fig.1 displayed in the upper subplot. 
 The various lines in Fig.2 for $p > 1$ correspond to the discrete spectrum of multipeak solutions.

The set of equations (viz. Eq.(\ref{d_cont}- \ref{wave})) 
are  numerically solved to study the time dependent problem. Equations (\ref{d_cont}) and 
(\ref{l_mom}) are evolved using the Flux Corrected Transport (FCT) algorithm \cite{boris76}. 
The vector potential as per the solutions discussed above is chosen to be circularly 
polarized. The wave equation (Eq.(\ref{wave})), which is a second 
order equation, is first written in terms of two coupled  first order convective equations 
which are then evolved by the FCT \cite{boris76} scheme. We choose periodic 
 boundary conditions along $x$. The box length is taken large enough to ensure that the excitations do not re-enter from the other boundary due to periodicity.  

In the next section we investigate the dynamical properties of the various  solitonic structures 
identified in the spectral diagram of Fig.2 obtained by Poornkala {\it et al.} \cite{kala1}  in a homogeneous 
 plasma background. For this we choose $n_0(x) = 1$.  
In section IV we consider an inhomogeneous background plasma density with a   
$ \tanh $ density profile.

\section{Soliton evolution studies in a homogeneous plasma}
We first evolve the  coupled set of equations Eq.(\ref{d_cont} - \ref{wave}) 
for an initial condition corresponding to the  small amplitude analytic $\sech$ soliton solutions 
obtained by Kaw {\it{et al.}} \cite{ksk} in the limit of weak density response. In this  limit and 
under the assumption of 
stationarity in a frame moving with velocity $\beta$ the coupled set of equations 
described in the last section,  reduce to a cubic nonlinear Schrodinger  equation which is known to  
admit exact soliton solutions. 
These solitonic structures are  stable and due to the infinite number  of invariants supported 
by the equations  preserve their identity even under intense nonlinear interactions suffered during  
collisions  with oppositely  propagating solitons. These features have been reproduced in our 
simulations (see Fig.3) and serve to
validate the correctness and accuracy of the code. 

We next carry out simulations for those solutions for which no analytic expression is available 
and whose forms  have been obtained numerically.  These structures are the general solutions 
with no restrictive assumptions  on the magnitude of the amplitude. As described in the last section,  
 these general nonlinear solutions 
have been well categorized with respect to their 
eigen - spectrum, group speed and number of wave cycles of  the vector potential 
structure in an earlier work \cite{kala1}. 
However, several questions pertaining to their dynamical properties remain open for investigation. 
Specifically the  issue of stability of these structures as well as the question of whether or not 
these structures possess solitonic traits of preserving individual identities in a 
collision process remain undetermined. To date there has been no theoretical investigation 
seeking an answer to these questions. 
Our approach in this paper has been to carry out a numerical study to shed light on this problem.
We find clear evidence that these  structures are stable 
for a very long time interval. However, the stability properties are different 
for the two classes of structures (viz. single and multiple  peak structures). The stability  seems to 
depend crucially on the number of peaks of the light wave field 
supported by the structure. The single peak solutions are very robust and have been 
observed to propagate undistorted for the entire duration of the simulation lasting for a 
few thousand plasma periods. 
However these robust single peak 
 solutions do not  preserve their identity under collisions - the characteristic feature 
 displayed  by   pure solitons. 
This shows  that these solutions are distinct from pure solitons and at best can  be regarded 
as solitary wave solutions. 
However, as has been the practice in the literature on laser plasma interaction,
 we will continue to refer to these structures as solitons. The distinction with exact solitons 
 will be made by refering to the value of their amplitudes (small for exact NLS solitons and 
 large for solitary waves solutions ) wherever necessary.


We first choose a particular  single peak high amplitude nonlinear solution depicted in the plot of Fig.4
corresponding to the following values of the parameters viz. $\lambda = 0.92$ and 
$\beta = 0.05$. Clearly for these values of $\lambda$ and $\beta$ the solution lies 
in the shaded area of the $\lambda - \beta$ plane in Fig.2.
For this solution  the weak density  approximation is not valid 
and hence it  does not satisfy the cubic 
nonlinear Schrodinger equation (NLS). In Fig.4 we display the difference in the structure of the exact solution 
with the sech expression (Eq.8 of Kaw {\it et al.} \cite{ksk}) obtained under the approximation of 
weak density response for the same values of parameters viz. $\lambda = 0.92$ and $\beta = 0.05$. 
It is clear from the figure that the difference between the two is quite significant and thereby it demonstrates the 
fact that these structures do not represent solutions in the weak density response regime.

We show in Fig.5 that this high amplitude solution propagates with a group velocity of $\beta = 0.05$ and 
survives for more than $1800$ plasma periods, (we have gone even beyond this time in our simulations   
 and have seen no perceivable distortion in its structure). This is a positive indication of 
 the stability of this solution. The evolution of a number of other large amplitude 
 numerically obtained single peak structures were investigated. They all propagate undistorted 
 during the entire simulation time which ranges upto a few thousands of  plasma periods. 
 
We now investigate the stability of other variety of structures namely those having  multiple light wave peaks. 
We show in Fig.6 
the amplitudes of the fields $\phi$ and $R$ and in Fig.7 the magnitude of the density at four different 
times in the evolution of one 
such multiple peak solution with $\lambda = 0.44466066$ and $\beta = 0.5 $. 
The structure does propagate  undistorted for several (almost $30$) plasma periods. However, it is not 
stable as one can see from the plots at later times where an additional structure appears to emerge 
from the trailing end. Basically, the multi peak structures are not stable 
and  emit radiation from their tail end. 
%
It should be noted here that the observed instability occuring  at the tail end can not be interpreted as 
though a multiple peak solutions is trying to break up into  stable single peak solutions. 
This is because the  amplitude of the exact single peak solitonic solutions  does not exceed  
the value of $R_m = \sqrt{3}$. (determined  from the restriction of  positivity of density for 
$\beta = 0$ \cite{esirkepov98}, 
for finite $\beta$ the amplitude is always less than this value, where it gets restricted by the lower 
curve of the shaded region in Fig.2). 

The very fact that the single peak solutions are stable and the mutiple peak solutions are not 
could be the reason that in several PIC simulations of the propagation of 
intense relativistic laser beam in plasma one only sees the formation of single peak solitonic structures 
\cite{bulanov99},  
and there are no reports on observations of multipeak  structures.


The next issue of interest  is to observe whether these solutions preserve their identity 
as and when they undergo collisions. We observe that the unstable multi peak structures 
cannot withstand any collisional interaction. Our numerical studies of mutual collisions between these 
solitons show that following a collision  
there remains no semblance to localized structures of any kind. Hence for  the study of collisional 
interaction we present results obtained for the robust single peak solutions only. 

To investigate  this we first choose two oppositely propagating 
large amplitude identical structures i.e. both having same $\lambda = 0.92$, but the group speed 
$\beta = 0.05$ for one and $\beta = -0.05$ for the other.  
In Fig.8 we depict the process of collisions between these two oppositely propagating structures. 
We observe that after the collision (see subplot for $t = 1800$) the emerging 
structures are identical to the incident ones. The only difference (which is more clear in the 
animated visualization) with Fig.3 (which depicts a collision between small amplitude solitons) 
is that in Fig.3 there occurs a time (e.g. $t = 1750$) of intense interaction 
where any semblance with the two impinging structure has dissappeared. The two structures basically 
spatially overlap. No such feature is observed in the collision depicted in Fig.8. As the structures 
approach each other, their group speeds  are observed to diminish and seem to vanish once the structures touch 
each other. At all times, for this particular  case of collision, 
amidst single peak, equal and large  amplitude structures, the two structures  continue to be 
identified separately. The structures then appear to reflect from each other.   
In this case, thus , even though 
the identity of the two structures after collision,  
seems to be preserved, however, the collisional interaction  seems to be  perceptibly  different from 
that exhibited by  pure solitons. However, we also observe that as 
 the value of $\mid \beta \mid $ for the two oppositely propagating equal amplitude structures 
is increased the collision 
 starts resembling that of low amplitude NLS solitons as can be seen from Fig.9, 
which shows the collisional interaction of a large amplitude solution with $\lambda = 0.9345$ and $\beta = 0.3$. 
This suggests that for low group speeds the collisional interaction seems to reflect the two 
structures, whereas at higher group speeds the two structures pass through each other undistorted.  
The two distinct varieties of collisions and their dependence of $\beta$ and amplitude would be discussed in detail 
in the next section where we try to understand them by interpreting the collisions as the propagation 
of one soliton on the inhomogeneous density of the other soliton.

However, on the basis of these 
studies alone one cannot arrive at any general conclusion 
on whether these large amplitude structures have solitonic traits or not. It is quite likely that 
the symmetry of the two underlying structures may have inhibited any exchange 
of energy between the two colliding structures, due to which they might have retained their identities 
even after they encounter each other. To pin  this down  we also carried out a study of 
collision amidst two large but different amplitude structures; their parameters being 
(e.g. one with $\lambda = 0.92$, $\beta = 0.05$ 
 and other 
with $\lambda = 0.87$ and $\beta = - 0.01$). The study of collision 
between these structures has been illustrated in Fig.10.  
In this case we observe that the identity of the two colliding structures gets lost after the interaction. 
 There is a clear exchange of energy amidst the two entities, as a result of which 
 the emerging structures after collision comprises of two peaked patterns, propagating in 
 opposite directions but having very different amplitudes than the original structures. 
 In fact it appears as though the intensity of the larger amplitude structure has increased 
 and that of the smaller one has diminished as they cross each other after colliding. 
 We observe that the larger amplitude structure formed after collision remains almost static. 
 These structures separately cannot be another set of solutions of the system from 
the same argument of  maximum 
 amplitude being   restricted to $R_m = \sqrt{3}$ \cite{esirkepov98} which is clearly exceeded 
 by the larger amplitude structure of Fig.10 at $t = 2600$.  
 There is also a significant amount of radiation (more apparant from the density plots of Fig.10) 
 which shows that not all the energy is localized in these structures.

>From the above studies  it becomes  clear that the destabilization (either inherent as in the multipeak 
structure or aided by the collisional interaction amongst two dissimilar  single peak solitonic structures ) 
predominantly leads to formation of long lived slowly varying  single peak structures which are 
either static or move  with low propagation speeds.  
The  amplitudes of these observed structures  are invariably higher than  
the value demanded  by the exact nonlinear solutions, hence they tend to radiate slowly. However, their survival 
for a considerably long time makes them an interesting set of transient structures.

\section{Soliton evolution studies in an inhomogeneous plasma}
In most experimental situations the background plasma medium has a density inhomogeneity. 
It is therefore a matter of practical interest to investigate the propagation of these large amplitude 
structures in an inhomogeneous medium. Here too we concentrate on studying the evolution 
of single peak structures through the inhomogeneous medium as they exhibit excellent stability 
characteristics. The inhomogeneity of the background plasma has been chosen to be 
of a simple tangent hyperbolic  form. 

We find that as the single peak structure  approaches a rarer  plasma background its speed 
increases. On the other hand if the background plasma  density  increases  the group speed is observed to decrease. 
We also observe that if the structures encounter a significant increase in the background plasma density, then 
the structures may even  get reflected from  a certain critical background density point. This has been shown in 
Fig.11 where we have plotted the amplitude  $R$ of the structure at various times with solid lines. 
The background plasma density has been taken to vary weakly with a $\tanh$ profile and the 
 dashed lines in the figure represent the background plasma density through 
which the structures propagate.
The five subplots in the  left column show the reflection of a large amplitude solution. For comparison 
we have also shown in the subplots of the right column the reflection of a small amplitude solution (which 
is an exact NLS soliton).  There is no  qualitative difference in the propagation features of 
large amplitude solutions and the small amplitude solutions of the NLS soliton variety,  
through the inhomogeneous plasma media. 

A detailed investigation of the dependence of critical background density $n_{ref}$ at which the 
structures reflect on parameters such as $\beta$, $\lambda$ and the amplitude $R_m$ has been made. 
This dependence has been depicted in the plot of Fig.12. We observe that $n_{ref}$ depends very weakly 
on $R_m$ and $\lambda$ but varies strongly with the group velocity $\beta$ of the structures. 
In fact it can be seen from the plot of $n_{ref}$ vs. $\beta$ shown in Fig.13 
that for any  value of $R_m$ (and/or $\lambda$ ) the points 
fit the curve  $n_{ref} = 1/(1-\beta^2)$ closely. 
The  dependence  $n_{ref} = 1/(1-\beta^2)$ corresponds to the 
dependence of critical density for reflection  on the group speed $\beta$ of linear wave packets. 
Basically one has from  the linear dispersion relation 
$\beta = (1-\omega_p^2/\omega^2)^{1/2} = (1-n_{00}/n_{ref})^{1/2}$ (here $n_{00} = 1$). 
The fact that this relationship is obeyed closely by these nonlinear structures is extremely interesting. 
It should be realised that  the amplitude of these structures are high only in a small 
central region. Though the  relativistic factor $\gamma$ at the maximum intensity  point for 
some of these solutions are around $1.3 - 1.4$ and the electron density gets evacuated from unity to 
around $0.8$ to $ 0.9$  causing the ratio to be as small as $n/\gamma \sim  0.6$ at the central region. 
Although this  suggests an effective  reduction of plasma frequency from unity to $0.7746$, 
but such a reduction happens only at the core region of the structure. At the edges, in fact 
an accumulation and hence enhancement of $n$ occurs and the intensity of the radiation field 
is also weak there. Clearly, if one employs the simplistic explanation of reflection being governed by local 
effective value of the plasma frequency, the edge of the soliton 
which is the first to encounter the inhomogeneous rise of density would reflect 
according to the linear relationship. This suggests that the nonlinear structure moves like a single coherent 
robust entity in the presence of weak inhomogeneity of infinite extent (e.g. such as the chosen 
$\tanh$ form) and instead of distorting or breaking merely follows the dynamics dictated 
by its edge which obeys the linear reflection relationship. It would be of interest to study 
whether the coherence of the nonlinear structure survives when it encounters a sudden significant 
rise in density for finite spatial region. In fact this question has been answered in our collision 
studies indirectly which we  discuss below. 
 
We first discuss the two distinct collisional behaviour observed in the context of equal amplitude solutions 
described in the last section in the light of above studies on reflection through inhomogeneous 
background plasma density. 
For $\beta = 0.05$ the critical density point for reflection would be  $n_{ref} = 1.0025$ from the 
relationship $n_{ref} = 1/(1-\beta^2)$. For the collision 
amidst two low amplitude NLS soliton moving with  this group speed  depicted in Fig.3 one can see that the 
value of maximum density of the structures are around $n_{max} = 1.001 < 1.0025$. Thus the 
density perturbation due to one soliton is insufficient to reflect the other. This results in strong overlapping 
of the two structures after which they emerge again moving in opposite directions. On the other hand it can be seen 
from Fig.8, where the large amplitude soliton moving with $\beta = 0.05$ is depicted, the value of maximum density 
due to  each of the structure is around $n_{max} = 1.0479 > 1.0025$. Thus the  individual structures get reflected 
as they encounter  the density inhomogeneity due to  the other structure.  In the previous  section 
we had also seen (Fig.9) that collision amidst two  large amplitude faster moving  $\mid \beta \mid = 0.3$ solitons, 
show traits similar to the low amplitude  NLS soliton. This can be explained as follows,  
here  $n_{ref} = 1/(1-\beta^2) = 1.0989$, whereas the $n_{max}$ of individual soliton (see $t = 0$) is 
much less than $1.01$ and even during intense interaction the perturbed 
density takes up a  maximum value of about $1.0721$. Both these values are less than $n_{ref}$, hence in this case 
instead of reflection a strong overlapping occurs and the two structures seem to pass through each other.

In these solitonic structures the electron density accumulation takes place only at the edge region 
(See Fig.1). 
The central region in fact gets evacuated and has lesser density. 
Thus the collision between two solitonic structures in some sense  mimics the propagation 
of one structure through a localized density inhomogeneity.  We take a relook at some 
of the collisions studied in the previous section with  this perspective and try to identify 
differences if any in the propagation through these two distinct classes of inhomogeneities 
(viz. sudden localized density inhomogeneity vs. slowly increasing inhomogeneity).  
The collision between unequal amplitude structures depicted in  Fig.10 points at an interesting difference. 
The energy exchange occuring in this case may be viewed as the smaller of the 
 two solitonic structures depicted in Fig.10 (coming from left) being unable to suffer a total  
reflection with its identity preserved (unlike  what  is observed of  $\tanh$ inhomogeneity )
as it encounters a 
 localized density inhomogeneity of the larger amplitude soliton coming from the right.  
After collision a very small amplitude remnant structure observed in the right side is the part 
which gets transmitted through the localized density inhomogeneity. A complete 
parametrization of reflection and transmission properties of these solitonic structures 
through localized inhomogeneities is in progress and would be presented in a subsequent publication soon.

\section{Summary and Discussion }
The present work deals with the study of  dynamical properties of one  
dimensional relativistic solitons in the form of modulated light pulses coupled to plasma waves. 
Earlier studies have focussed primarily on obtaining various traveling wave solutions and their categorization 
in terms of their group speeds, frequency, and the form of solution (e.g. number of light wave 
peaks in the solution etc.). It was also shown earlier that the small amplitude solutions 
obtained under the approximation of weak density response were identical to the NLS solitons. 
However, nothing conclusive as regards the stability and possibility of exhibiting 
any kind of solitonic traits was known  about  the large amplitude solutions. We have used the 
combined fluid-Maxwell set of equations to evolve such large amplitude solutions in time to 
ascertain some of these properties. 

Our simulations show that the large amplitude solutions in which there is only one 
peak of trapped light wave are very robust and survive for the entire (several thousands 
of plasma periods) simulation duration.  The structures which have multiple peaks of light waves 
enclosed within the modulated structure 
are seen to survive in some cases only upto a few plasma periods (in the example shown it is upto $30$ 
plasma periods). However, these  structures are found to be susceptible to 
an instability which develops at their trailing edge. As a result of this instability these 
structures are extremely fragile. 

The single peak large amplitude structures though extremely stable are in some sense 
distinct from  true solitons.  They do not 
preserve their identities upon collisions. Only under the special case when 
the oppositely propagating structures are identical that  they  
preserve their identity after collisional interaction. In all other collisional encounters 
there is always an exchange of energy amidst the two colliding structures and also 
 a significant transfer of energy into radiation.  Hence these structures can at best be regarded 
 as solitary waves and are not true solitons which are constrained by infinite number of conservation laws. 

When the colliding structures are identical they display two distinct variety of interactions 
depending on the group speed $\beta$ and the amplitude of the structures. 
In one case the structures merely reflect from each other (the overlapping is weak 
and  in fact negligible),  while  in the other case the structures approach each other,
overlap and seem to pass through each other. These features and their dependence on $\beta$ as well as  on their amplitude 
can be understood by realising that the solitons as they approach each other view the density enhancement at the edge 
 of the other structure as an  inhomogeneity. 

The propagation of these structures through background inhomogeneous plasma density has also been studied. 
These studies show qualitative resemblance with the features displayed by exact NLS  solitons 
as they pass through density 
inhomogeneity. For instance the group speed of these structures decreases (increases) as the background density 
increases (decreases).  They also suffer total reflection  from weakly 
varying density inhomogeneity as an exact NLS soliton would do.
We have studied in detail parametric dependence of the  critical density for reflection $n_{ref}$. 
The studies reveals interestingly a very weak dependence on the amplitude $R_m$
and the eigen value $\lambda$ of the solitonic structure. However, a strong dependence on $\beta$ of the form 
$n_{ref} = 1/(1-\beta^2)$ has been observed. This dependence on group speed $\beta$ is identical to  what a linear 
wave packet does in the presence of density inhomogeneity. This is  interesting as it shows that 
even though the structure is strongly nonlinear, there is negligible manifestation of this nonlinearity 
in the reflection property. 

The quantitative dependence of the form $n_{ref} = 1/(1-\beta^2)$ helps in identifying 
the parameter regime for the two varieties of collisions occuring for equal amplitude solitonic structures 
described in the paragraph above. Reflection occurs whenever $1/(1-\beta^2) <  n_{max}$ (here $\mid \beta\mid$ 
is the group speed and $n_{max}$ is the maximum amplitude of  density of each of the identical colliding structures), 
and a strong overlapping takes place when the inequality is reverse. 

Our studies on collision between unequal amplitude structures also indicates  that 
propagation through localized inhomogeneity   may show distinctive features. A part of  soliton can get 
transmitted through a localized inhomogeneity provided the width is small and amplitude is also 
appropriate. A detailed investigation is underway and would be presented elsewhere soon.   
 
The present work has thus shown conclusively that 
while most of the evolution characteristics of single peak large amplitude structure resemble 
soliton solutions, they still do not qualify to be exact solitons in a strict mathematical sense. 
However, their robustness and their  close resemblance to 
solitonic properties can make them  quite useful in practical applications such as 
stable transfer of energy packets deep into laser fusion targets or coherent propagation in plasma based accleration
devices. \\

\newpage 
\begin{center}
{\bf FIGURE CAPTIONS}\\
\end{center}

\noindent
FIG. 1. The spatial profile of $R$ (solid lines), $\phi$ (dotted lines) and $n$ (dashed lines) for single 
peak (upper subplot) and multiple peak (lower subplot) 
 nonlinear travelling wave solutions for the coupled laser plasma system. The upper subplot is for 
$\lambda = 0.92$ and $\beta = 0.05$. The multiple peak structure 
corresponds to $\lambda = 0.44466066$ and $\beta = 0.5$. \\

\noindent 
FIG. 2.  The  $\lambda - \beta$  values for which single peak (shaded region) and multiple peaks 
(the lines with various $p$ values; with $p$ indicating the number of extrema in $R$) are possible.\\

\noindent
FIG. 3. Shows the collisional interaction amongst 
two small amplitude single peak solitons. The parameters corresponding to 
the two structures are   
 $\beta = \pm 0.05$ and $\lambda = 0.944 $. The plot of $R$, $\phi$ and $n$ representing 
vector potential, scalar potential and electron density respectively at three different times are shown. 
The $t = 0$ plot shows the initial condition, at $t = 1750$ the interaction is maximum 
and $t = 3600$ shows the configuration after collision. Here time and lengths are 
normalized to $\omega_{p}^{-1}$ and $c \omega_{p}^{-1}$ respectively.\\

\noindent
FIG. 4. Comparison of high amplitude solution obtained numerically (dashed line) for 
$\beta = 0.05$ and  $\lambda = 0.92$ with  sech profile for these parameters 
(solid line) obtained from  Eq.8 of Kaw {\it et al.} \cite{ksk}. The difference clearly 
shows that for these solutions the approximation of weak density response is not valid.\\

\noindent
FIG. 5. Shows the time evolution of a high amplitude single peak solutions  having 
$\beta = 0.05$ and $\lambda = 0.92$.  The solid, dashed and dot dashed plots represent 
the structure at $t =0$, $t = 900$ and $t = 1800$ respectively. The figure clearly shows 
that these solutions are stable. \\

\noindent
FIG. 6. Shows  plots of $R$ (solid line) and $\phi$ (dashed line) profiles for a
 multipeak structure at four different times, here $\beta =0.5$ and $\lambda = 0.44466066$. 
 The structure retains 
 its identity upto about $30$ plasma periods after which it starts shedding radiation 
 from the tail end. \\

\noindent
FIG. 7. The four subplots show the profile of  the  electron density corresponding 
to the data and subplots of Fig.6. \\

\noindent
FIG. 8. Shows the results of collisional interaction amidst two 
oppositely propagating high amplitude single peak identical solutions. The parameters for the two 
are $\beta = \pm 0.05$ and $\lambda = 0.92 $. The two structures seem to  reflect.\\

\noindent
FIG. 9. Shows the results of collisional interaction amidst two 
oppositely propagating high amplitude single peak identical solutions. The parameters for the two 
are $\beta = \pm 0.3$ and $\lambda = 0.9345 $. Here the group speed is much higher than that of Fig.8. 
The structures show strong overlapping and seem to pass through each other. \\

\noindent
FIG. 10. Collisional interaction amidst two high amplitude single peak structures which are 
not identical.  One
 (at $t = 0$ on left) has  $\beta =0.05$ and $\lambda = 0.92 $ and other 
(at $t = 0$ on right) with  $\beta = -0.01$ and $\lambda = 0.87 $. It 
is clear (compare  plots at $t = 0$ with  $t = 2600$) that the structures do not retain 
their identity after collision. \\

\noindent 
FIG. 11. Shows the comparison of reflection from inhomogeneous density 
of the high amplitude single peak solutions (subplots in left column) with that of a small 
amplitude NLS soliton (subplots in right column).The background density has been 
illustrated by dashed lines and its values are depicted by the vertical axis on the right side 
of each subplots. \\

\noindent
FIG. 12. Plot showing the parametric dependence of $n_{ref}$ (the value 
of background density at which reflection occurs) on the soliton amplitude $\lambda$ (upper subplot)
 and $R_m$ (lower subplot) for various $\beta$. The plots suggests that the dependence on $R_m$
and $\lambda$ is weak. In the inset we have shown this weak dependence by magnifying the axis for 
$\beta = 0.1$. \\

\noindent
FIG. 13. Variation of the critical background density  $n_{ref}$ at which reflection 
occurs with the group speed for the entire range of $R_m$ and $\lambda$ depicted in the data 
of Fig.12. The circles are obtained from simulation and the 
dashed curve shows the fit to $1/(1-\beta^2) $. \\

\end{document}